\documentclass{ws-procs975x65}

\newcommand{\e}{\epsilon}

\newcommand{\tp}{t^\prime}
\newcommand{\ep}{\epsilon^\prime}
\newcommand{\mup}{\mu^\prime}

\newcommand{\lsim}{\lower.5ex\hbox{$\; \buildrel < \over\sim \;$}}
\newcommand{\gsim}{\lower.5ex\hbox{$\; \buildrel > \over\sim \;$}}

\begin{document}

\title{Photoabsorption of Gamma Rays in Relativistic Jets}

\author{Charles D.\ Dermer}

\address{Code 7653, Naval Research Laboratory\\
4555 Overlook Ave.\ SW,  
Washington, DC 20375-5352 USA\\ 
E-mail: dermer@gamma.nrl.navy.mil}

\maketitle

\abstracts{
A derivation of the $\gamma\gamma\rightarrow$ e$^+$ e$^-$ optical
depth for $\gamma$ rays produced in a comoving spherical emitting
region is presented.  Employing a simplified expression for the
$\gamma\gamma$ absorption cross section, analytic expressions for the
minimum Doppler factor implied by the requirement of $\gamma$-ray
transparency are derived for a broken power-law spectrum of target
photons which are isotropically distributed in the comoving
frame. Application to specific systems is illustrated.  }

\section{Introduction}

One particularly powerful probe of relativistic motions of AGN and GRB
jets, as revealed by the EGRET instrument on the {\it Compton
Observatory}, is the use of $\gamma$-ray observations to infer minimum
Doppler factors of the radiating plasma. The basic idea is simple: the
measured FWHM variability time scale $t_{var}$ of a blazar flare or
GRB pulse implies a maximum comoving radius of the emitting region
from causality considerations. The measured flux and redshift implies
the corresponding density of photons which provide targets for
$\gamma\gamma\rightarrow$ e$^+$ e$^-$ pair production attenuation of
$\gamma$ rays.  The requirement that the emitting region have small
optical depth for observed $\gamma$ rays places a lower limit on the
Doppler factor $\delta$ of the emitting region. In many cases, such
arguments indicate that the radiating plasma in blazar and GRB jets
must be relativistic\cite{dg95,ls01}.

The dramatic improvements in sensitivity of the upcoming {\it GLAST}
mission and the ground-based imaging air Cherenkov telescopes {\it
VERITAS} and {\it HESS} over previous instruments offer the opportunity
to place better limits on $\delta$, to monitor changes of $\delta$ in
a given source, and to compare $\delta$ between members of different
source classes, e.g., BL Lac objects and flat spectrum radio quasars
(FSRQs), and X-ray rich, short duration, and classical GRBs. Because
{\it GLAST} is a scanning mission, blazar flares can be correlated
with radio outflows to infer the locations of the sites of
$\gamma$-ray emission.  The $\gamma$-ray observations, coupled with
correlated multifrequency data, will provide knowledge of the jet-disk
connection, jet dynamics, and radiation fields in the vicinity of the
jet. For these reasons, it seems appropriate to revisit the problem of
$\gamma$-ray photoabsorption in relativistic jets.

\section{Derivation of the Minimum Doppler Factor}

We consider a uniform spherical radiating blob of volume $V^\prime_b =
4\pi r_b^{\prime 3}/3$, where primes refer to the comoving frame.  The
measured $\nu F_\nu$ flux is related to the comoving emissivity
according to the relation
\begin{equation}
\nu F_\nu = f_\epsilon \cong {\delta^4 V^\prime_b\over d_L^2}
\;\epsilon^\prime j^\prime(\ep ;\vec\Omega^\prime)\rightarrow 
{\delta^4 V^\prime_b\over 4\pi d_L^2}\;\epsilon^\prime j^\prime(\ep )\; ,
\label{vFv}
\end{equation}
where $j^\prime(\ep , \vec\Omega^\prime)$ is the comoving emissivity,
$d_L=10^{28}d_{28}$ cm is the luminosity distance, $\e = h\nu/m_e c^2
= \delta \ep/(1+z)$ is the dimensionless photon energy, $z$ is the
source redshift, $\vec\Omega$ is the direction vector, and the final
expression in eq.\ (\ref{vFv}) assumes isotropic emission in the
comoving frame.  Eq.\ (\ref{vFv}) is an approximation because we have
implicity assumed that the light travel time across the blob is
smaller than the duration of the flare in order to avoid integrations
over different portions of the emitting plasma, consistent with
causality arguments for a discrete intense flare or pulse.  The
expression $dt = (1+z) dt^\prime/\delta$ relating observer and
comoving differential times implies $r_b^\prime \lsim c
t^\prime_{var} = c\delta t_{var}/(1+z)$.

The comoving spectral energy density $u^\prime_{\epsilon^\prime} =
 m_ec^2 \e^{\prime 2}n^\prime (\ep )$ is related to the isotropic
 emissivity $j^\prime(\ep)$ through the relation $\ep j^\prime(\ep)
 \cong u^\prime_{\epsilon^\prime}/\langle
\tp_{esc}\rangle \cong cu^\prime_{\epsilon^\prime}/r_b^\prime$, where
$\langle \tp_{esc}\rangle$ is the mean photon escape time and
$n(\ep)$ is the spectral photon number density. Thus
\begin{equation}
u^\prime_{\epsilon^\prime}
\cong {3 d_L^2\over c r_b^{\prime 2} \delta^4}f_\e = 
{3 d_L^2\over c r_b^{\prime 2} 
\delta^4}f_{\e_{pk}}[x^a H(1-x) +x^b H(x-1)]\;,
\label{uprime}
\end{equation}
where $x = \e/\e_{pk} = \ep/\ep_{pk}$, $H(u)$ is a Heaviside function,
and we have approximated the $\nu F_\nu$ spectrum by a broken power
law with peak energy flux $f_{\e_{pk}}$ at $\e = \e_{pk}$ with $\nu
F_\nu$ indices $a~ (> 0)$ and $b~ (< 0)$ at low and high frequencies,
respectively.

The photoabsorption optical depth for a $\gamma$-ray photon with
energy $\e_1$ in a radiation field with spectral photon density
$n(\ep ,\mup ;r^\prime)(\approx n(\ep)/2$ for a uniform isotropic radiation
field in the comoving frame) is\cite{gs67}
$$\tau_{\gamma\gamma}(\ep_1) =
\int_{r_1^\prime}^{r_2^\prime} dr^\prime \int_{-1}^1 d\mup (1-\mup)
\int_{2/\ep_1(1-\mup )}^\infty d\e\; \sigma_{\gamma\gamma}
[\ep\ep_1(1-\mup)]n(\ep ,
\mup ;r^\prime)$$
\begin{equation}
\cong r_b^\prime \int_0^\infty d\ep \; \sigma_{\gamma\gamma }
(\ep,\ep_1)n(\ep ).
\label{taugg}
\end{equation}
 
Using the approximation\cite{zl85} $\sigma_{\gamma\gamma }
(\ep,\ep_1)\cong {1\over 3} \sigma_{\rm T} \ep
\delta(\ep - 2/\ep_1)$ for the cross section and
making the substitutions in eq.\ (\ref{taugg}) gives
\begin{equation}
\tau_{\gamma\gamma}(\e_1) = \tau_{\gamma\gamma}^{pk}\;
[({\e_1\over \e_1^{pk}})^{1-b}H(\e_1^{pk}-\e_1) + 
({\e_1\over \e_1^{pk}})^{1-a}H(\e_1-\e_1^{pk})]\;,
\label{tauggap}
\end{equation}
where
\begin{equation}
\tau_{\gamma\gamma}^{pk}\cong {\sigma_{\rm T} 
d_L^2 f_{\e_{pk}}\over 4 m_ec^4 t_{var} \delta^4 
\e_{pk}}\;\;{\rm and} \;\; \e_1^{pk} = 
{2\delta^2\over (1+z)^2 \e_{pk}}\;.
\label{tgge1}
\end{equation}
Requiring that $\tau_{\gamma\gamma}(\e_1)< 1$ so that the emission
region is transparent to $\gamma$ rays, we obtain a minimum Doppler
factor given by
\begin{equation}
\delta > [ {\sigma_{\rm T} d_L^2 f_{\e_{pk}}\over 
4 m_ec^4 t_{var} \e_{pk}^A}\bigl({(1+z)^2 \e_1\over 2}
\bigr)^{1-A}\;]^{1/(6-2A)}\;.
\label{delta}
\end{equation}
In eq.\ (\ref{delta}), $A =b$ if $\e_1 < \e_1^{pk}$, and $A = a$ if
$\e_1> \e_1^{pk}$ (note that $A = 1-\alpha$ where the energy index $\alpha$
is defined by the relation $F_\nu\propto \nu^{-\alpha}$).

\section{Application to Observations}

We first apply eqs.\ (\ref{tgge1}) and (\ref{delta}) to FSRQs such as
3C 279, PKS 0528+134, or CTA 102.  OSSE and COMPTEL
observations\cite{mcb95} of bright blazars show that $f_{\e_{pk}}=
10^{-10}f_{-10}$ ergs cm$^{-2}$ s$^{-1}$ with $f_{-10}\gsim 1$, and
$f_{-10}\approx 10$ during bright blazar flares.  The $\nu F_\nu$ flux
peaks at $\e_{pk}
\sim$ 1--100. Photons observed at 100 MeV/GeV energies (i.e., $\e_1
=1960 E_{GeV}\gsim$ 200--2000) would therefore be preferentially
absorbed by photons with energies below the peak of the $\nu F_\nu$
spectrum, so that the case with $A=a$ applies, and $a \sim $
0.2--1. Taking $a = 0.5$ for illustration gives
\begin{equation}
\delta \gsim 6.8 \; \bigl[ {d_{28}^2 f_{-10}
\over t_4}\;\bigl({1+z\over 2}\bigr)\bigr]^{1/5}
\;\bigl( {E_{GeV}\over \e_{pk}}\bigr)^{1/10}\;.
\label{dmin}
\end{equation}
Here we have used $t_4 = t_{var}({\rm s})/10^4 {\rm ~s}\sim 1$ as a
fiducial because {\it GLAST} will be able to detect variability from
flares at the level of $f_{-10} \sim 1$ on this timescale\cite{dd03}.
For bright blazar flares from distant FSRQs, {\it GLAST} could set
minimum values of $\delta$ exceeding $\sim 20$, comparable to the
largest values inferred from radio observations of superluminal
motion.

We next consider TeV blazars ($\e_1 = 1.96\times 10^6 E_{TeV}$) such
as Mrk 421 or Mrk 501 at a distance of $\approx 140$ Mpc. TeV photons
from sources with $\delta \approx 10$ preferentially pair produce with
$\sim 1$ keV synchrotron photons, the X-ray synchrotron spectrum can
exceed $f_{-10} \approx 1$, and the flare timescale $t_4$ may be as
low as $0.1$, or even lower, depending on results from the next
generation of ground-based $\gamma$-ray telescopes. Using $A = a = b =
0$ for simplicity (a flat $\nu F_\nu$ synchrotron spectrum) gives
\begin{equation}
\delta \gsim 10.9 \; \bigl[ \bigl({d_L\over 140{\rm ~Mpc}}\bigr )^2
{f_{-10} E_{TeV}
\over t_4}\bigr ]^{1/6}\;.
\label{dmintev}
\end{equation}
Shorter flares and brighter synchrotron fluxes will potentially imply
values of $\delta \gsim 15$.

Our results can also be applied to GRBs. The main difference here is
that the geometry of emission for GRBs is generally considered to be a
spherical blast wave that is uniform within the Doppler beaming
cone. This geometry for GRB emission lacks strong observational
support, so a blob geometry might equally well apply to a GRB; in any
case, the differences between blast-wave and blob geometries do not
make a great deal of difference in the resultant expressions for the
pair-production opacities (Dermer 2004, in preparation).

The prompt hard X-ray/soft $\gamma$-ray emission of a GRB, which we
assume here to be nonthermal synchrotron radiation, has $\e_{pk} \sim
1$. GRBs with peak flux $f_{\e_{pk}}= 10^{-6} f_{-6}$ ergs cm$^{-2}$
s$^{-1}$, with $f_{-6}\gsim 1$, occur every 2 -- 4 weeks over the
full sky.  Anticipating that $\delta_{100}=\delta/100 \approx 1$ for
GRBs, we see that $\e_1^{pk}
\approx 5000\delta_{100}^2/\{[(1+z)/2]^2\e_{pk}\}$, so that
observations of GeV photons from GRBs generally favor the case $\e_1
\lsim \e_1^{pk}$. In other words,
 GeV photons from GRBs are preferentially attenuated by photons on the
high-energy portion of the synchrotron spectrum. For illustration, we
let $A = b = -1/2$ for a typical value of the spectral index at
$\gsim$ MeV energies, and assume that the synchrotron spectrum extends to
GeV energies, giving
\begin{equation}
\delta \gsim 177 \; \bigl[ {f_{-6}d_{28}^2
\over t_{var}({\rm s})}  
\bigl({1+z\over 2}\bigr)^3\bigr]^{1/7}
\;\bigl [ \e_{pk} E_{GeV}^3 \bigr]^{1/14}.
\label{dminGRB}
\end{equation}
Thus we can expect {\it GLAST} observations of GRBs to infer values of
$\delta\gg 200$ in the brightest and most variable GRBs, and
potentially infer differences in values of $\delta$ between
``classical" GRBs and X-ray flashes, which are predicted to be low
$\delta$-factor GRBs\cite{dcb99}.

\section{Discussion}

The results here are in essential agreement with previous treatments,
taking into account the different approximations made for the cross
sections\cite{dg95,ls01}. Note an implicit co-spatial assumption in
the derivation, namely that the $\gamma$-rays are formed in the same
region as the lower energy target photons. Without this assumption,
only much smaller values of minimum Doppler factors can be confidently
asserted, which depend more on observations at MeV energies rather
than at GeV or TeV energies\cite{dg95a}. To demonstrate the
reliability of the co-spatial assumption, and to measure the spectrum
of photons that attenuate the $\gamma$ rays, {\it GLAST} and
ground-based $\gamma$-ray observations of blazars and GRBs should
be correlated with observations made by X-ray/soft $\gamma$-ray
detectors such as {\it INTEGRAL}, {\it Chandra}, {\it XMM Newton} and
{\it Swift}.

\section*{Acknowledgments}
I thank Drs.\ Roberto Turolla and Silvia Zane for the opportunity to
speak in the session on Radiative Transfer in Relativistic
Astrophysics at the 10th Marcel Grossman Meeting. This work is
supported by the Office of Naval Research and {\it GLAST} Science
Investigation Grant No.\ DPR-S-1563-Y.

\end{document}